\documentclass[dvips]{article}

\usepackage{icrc2011}

\title{Do we observe fluctuation of cross section in cosmic rays ?}

\newcommand{\etal}{\MakeLowercase{\textit{et al. }}} 
\shorttitle{G. Wilk \etal Do we observe cross section fluctuations
...}

\authors{Grzegorz Wilk$^{1}$, Zbigniew W\l odarczyk$^{2}$}
\afiliations{$^1$The Andrzej So\l tan Institute for Nuclear
Studies, Ho\.za 69, 00-681 Warsaw, Poland\\ $^2$Institute of
Physics,Jan Kochanowski University, \'Swi\c{e}tokrzyska 15, 25-406
Kielce, Poland}
\email{wilk@fuw.edu.pl}

\abstract{It is known from some time that to explain some
apparently {\it exotic events} observed in cosmic ray experiments,
a deeply penetrating component in cosmic rays is needed, which, in
turn, leads frequently to the nonexponential distributions of some
observables. We argue that such unexpected behavior of some cosmic
rays data can be most naturally explained by attributing them to
the possible fluctuations of cross sections.}

\keywords{ The keywords will be used to select your subject from all
ICRC contributions. }

\begin{document}
\maketitle

\section{Introduction}

Since quite a time already, a number of existing experimental
results obtained by the Emulsion Chamber Experiments can be
regarded as {\it exotic}. They were so far not encountered in the
accelerator experiments and, notwithstanding developments in
modern understanding of the elementary particle phenomena, they
are remaining somehow forgotten and still wait for their proper
understanding \cite{ZW_Calgary,LFC1,LFC2,Lattes}. Those are, for
example, long flying components, centauro species or coplanar
emission phenomenon. In what follows we shall very shortly remind
properties of these chosen examples and we shall propose a common
possible explanation: a possible fluctuation of the cross section
remaining behind them in one or other way. Our aim is to bring
ones attention to the fact that there is quite a number of
unexplained phenomena in CR which still await their consideration,
especially at present, with new experiments at LHC making contact
with the CR experiments\footnote{One should keep always in mind
that this can be really at most contact, not an overlap. The
reason is that CR experiments probe the forward region of the
phase space whereas accelerators are mostly concentrated in its
central part.}.

A word of justification in what concerns the use of notion of
fluctuating cross section is in order here. It was introduced and
discussed in \cite{csfluct1,csfluct2} (to describe results of some
diffraction dissociation experiments on accelerators) and it was
again invoke recently when discussing some Hera and Tevatron
\cite{csfHera} or LHC \cite{csfLHC} data. It means that this
subject is still alive and worth of consideration. In CR it was
used for the first time long time ago by us (cf. remarks below)
\cite{WW_PRD}).

\section{Some examples of {\it exotic events}}

We shall now remind shortly the main features of the chosen {\it
exotic events}. Only essential features will be presented, all
details can be found in the references provided.

\subsection{Long flying component (LFC)}

This name was coined for the phenomenon observed when
investigating the propagation of the initial flux of incoming
nucleons \cite{LFC1,LFC2}. They can be presented in different ways
but for the purpose of this presentation we shall concentrate only
on the example of the distribution of the cascade starting points
in thick lead chamber of the Pamir experiment \cite{LFC1}. What is
regarded as peculiar, or even unexpected, is the fact that the
commonly used exponential formula
\begin{equation}
\frac{dN}{dt} = f(t) = \frac{1}{\lambda} \exp\left( -
\frac{t}{\lambda }\right) \label{eq:exp}
\end{equation}
is definitely invalid (especially at large depths $t$, $\lambda
\sim 1/\sigma$ is the mean free path and $\sigma$ is the
corresponding cross section). The observed discrepancy look like
hadrons tend to fly longer without interaction (therefore the term
{\it long flying component} for this type of phenomenon). This was
the place where we have used, for the first time in CR, the idea
of fluctuating cross section \cite{WW_PRD}. It turns out that with
the assumed variance of the cross section equal to $\sim 0.2$ one
can nicely reproduce the observed enhanced tail of the
distribution of the cascade starting points. It should be also
stressed that the idea of fluctuations invoked here was prompted
us, as explained below, towards applications of the so called
Tsallis distribution (and, in general, Tsallis statistics)
\cite{T} to describe power-like behavior of many distributions
observed both in CR \cite{WW_NP} and in the usual multiparticle
production phenomena \cite{WW_epja}.

\subsection{Centauros}

The other phenomena requiring a kind of deeply penetrating
component are the so called {\it Centauro's} and {\it
mini-Centauro's}. Their characteristic feature is the extreme
imbalance between hadronic and gamma-ray components among the
produced secondaries. Actually, they are the best known example of
numerous unusual events reported in cosmic-ray experiments
\cite{Lattes}. There are many attempts to explain them, like
different types of isospin fluctuations or formation of
disoriented chiral condensate (DCC) \cite{DCC}, multiparticle
Bose-Einstein correlations (BEC) \cite{BE}, strange quark matter
(SQM) formation \cite{SQM,ZW_NP52}). All of them reproduce many
features of Centauros in a single collision but fail to explain
the substantial number of interactions contributing to the
development of families observed at mountain altitudes among which
Centauro were observed. Actually, it was shown that families
recorded at mountain altitudes are insensitive to any isospin
fluctuations \cite{ZW_NP52, ZW_JPG19}.

\subsection{Coplanar emission}

Phenomenon of alignment of structural objects of gamma-hadron
families near a straight line in the plane at the target diagram
was first observed during examination of multicore halos and,
later, when observing distinguished energetic cores (i.e. halos,
energetic hadrons, high energy gamma quanta or narrow particle
groups) \cite{Kopenkin,Borisov,Antoni}. The excess of aligned
families found in these cascades exceed any known conventional
concept of interaction. Many attempts to interpret this phenomenon
of coplanar emission were undertaken. For example, in \cite{HM} it
was proposed to be due to the copious occurrence of semihard jets
(with $p_t > 3$ GeV). However, such interpretation is not
satisfactory because of the noticeable difference in the energy
distribution of the main streams of secondaries (jet particles
have too low energy). In \cite{R} coplanar phenomena were
attributed to the projection of the quark-gluon string ruptures
produced in the process of semi-hard double inelastic diffraction
dissociation (with string being inclined between a semi-hard
scattered fast quark and the incident hadron remnants). This
explanation seems plausible because the energy threshold of the
alignment effect is consistent with the threshold-like dependence
of semi-hard double inelastic diffraction. The "length" of aligned
groups of EDC as a projection is also more or less in agreement
with the transferred momentum while string production . In this
case the target diagram of a superfamily with alignment may be
considered as a direct "photographic" image of such. However,
until now no truly satisfactory explanation exists, especially
when realizing that the observed alignment occurs not so much in
the elementary interaction but during the development of the full
cascade, which is a macroscopic process with substantial number of
interactions contributing to family formation. Also here the
long-flying component with the mean free path of the order of the
few hundreds of $ g/cm^{2} $ is required \cite{Borisov}.

\section{Nonexponential behavior}

The standard modelling used to analyze cosmic ray data assumes the
constant cross section at given energy. However, since some time
already the evidence is accumulating that such approach is too
simplified because it does not account for the possible intrinsic
fluctuations in cross section which result in a characteristic
power-like behavior of the depth distribution of the starting
points of cascades \cite{WW_NP}. The origin of the cross-section
fluctuations (CSF) can be traced down to the fact that hadrons
have internal degrees of freedom (color-carrying quarks and
gluons) and can therefore collide in different internal
configurations resulting in different cross sections. Fluctuations
of the hadronic cross sections are discussed in the literature and
observed in diffraction dissociation experiments on accelerators
\cite{csfluct1,csfluct2,csfHera,csfLHC}. We refer for details and
physical justification of this phenomenon to
\cite{csfluct1,csfHera}.

Actually, as was shown in \cite{WW}, such fluctuations inevitably
lead to the so called Tsallis statistics \cite{T}, which in a
natural way enlarges the predictive power of the usual statistical
approaches. For completeness, we list some basic features of
Tsallis statistics \cite{T}. It is well known \cite{WW,B,Biro,
fluct, WW_epja} that whenever in the exponential
 formula (\ref{eq:exp}) the parameter $\lambda  \propto1/\sigma$
 fluctuates according to gamma distribution,
\begin{eqnarray}
g(1/\lambda) &=& \frac{1}{\Gamma\left(\frac{1}{q - 1} -
s\right)}\frac{\lambda_{0}}{q - 1}\left(\frac{1}{q -
1}\frac{\lambda_{0}}{\lambda}\right)^{\frac{1}{q - 1} - 1 - s}\cdot \nonumber\\
&&\cdot \exp\left(\frac{1}{q -
1}\frac{\lambda_{0}}{\lambda}\right) \label{eq:gamma}
\end{eqnarray}
one obtains as a result the power-like distribution
\begin{eqnarray}
h_q(t) &=& \int_0^{\infty}f(t)g(1/\lambda)d(1/\lambda) =
\nonumber\\
&=& \frac{2-q}{\lambda_{0}}\left[ 1 - (1 - q)\frac{t}{\lambda_{0}}
\right]^{\frac{1}{1 - q}}. \label{eq:qexp}
\end{eqnarray}
Here $\lambda_{0}$ denotes the value of $\lambda$ around which one
has fluctuations given by Eq. (\ref{eq:gamma}) and parameter $s$
indicates to which classes of superstatistics (notion introduced
in \cite{supers}), type A with $s=0$ or type B with $s=1$, this
distribution belongs (we put $ s=0 $ in all analysis of cosmic ray
data). In both cases the so called nonextensivity parameter $q
\in(1,2)$ reflects the amount of fluctuations with
\begin{equation}
q = 1 + \frac{\omega}{1 + s\cdot\omega},\quad  {\rm where}\quad
\omega = \frac {\langle \lambda^{2} \rangle - \langle \lambda
\rangle^{2} }{\langle \lambda\rangle^{2}}.\label{eq:Q}
\end{equation}
Notice that for $q \rightarrow 1$ Eq. (\ref{eq:qexp}) becomes Eq.
(\ref{eq:exp}) used before.

Available experimental data at low energies (up to $ 10^{4} $ GeV
in lab.) suggest that  relative variance of cross section increase
with energy as $ \omega\sim \ln s $ or is asymptotically constant
$ \omega\sim b-c/s^{\varepsilon} $ \cite{csfluct1}. As mentioned
above, in Ref. \cite{WW_PRD} distribution of depths of starting
points of cascades in Pamir lead chamber was described by Monte
Carlo simulation using the cross section sampled from the uniform
distribution, $$ \sigma\in
[(1-\sqrt{3\omega}\sigma_{0}),(1+\sqrt{3\omega}\sigma_{0})],$$
with $ \omega\simeq 0.2 $. It was then proposed in \cite{WW_NP} to
resort to Tsallis distribution (\ref{eq:qexp}) (corresponding to
fluctuations of the cross section according to the gamma
distribution (\ref{eq:gamma})) to describe the same Emulsion
Chamber data and the relative fluctuations of cross section found
there was $ \omega\simeq 0.3 $.

\section{Concluding remarks}

Apart of the above scenario (i.e., CSF) aiming to explain {\it
long flying component}, the other hypothesis have been widely
discussed in the literature \cite{LFC1, LFC2}:
\begin{itemize}
\item [$(i)$] charm particles (charmed $ \Lambda_{c} $ hyperons
and $ D $ mesons) with $ \sigma_{charm-Pb}<\sigma_{p-Pb} $ are
produced somewhere in the upper parts of chamber;

\begin{figure}[h]
  \centering
  \includegraphics[width=3.3in]{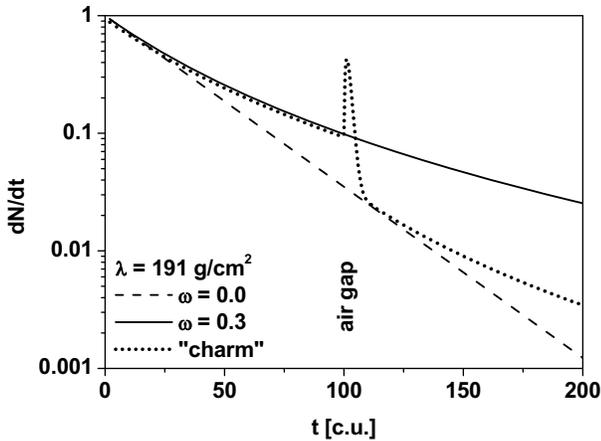}
  \caption{Schematic depth distribution of the starting points of cascades
           in the two-storey Pb chamber. The dotted line shows absorption curve
           in the case of production of charmed particles. Full line
           shows results for fluctuating cross section. The normal attenuation
           are shown by broken curve. All distributions are normalized to
           $1$ at $t=0$.
    }
  \label{fig1}
 \end{figure}

\item[$(ii)$] there exist some new weakly absorbed hadrons not yet
found in accelerators.
\end{itemize}
The weakness of the first hypothesis is that in order to explain
the observed effect  we need to assume that the production cross
section of charmed particles is as high as $
\sigma_{\Lambda_{c},D}\simeq  $ 2 mb/nucleon, while extrapolation
of accelerator data gives one order less value for $
\sigma_{\Lambda_{c},D} $. In order to check different hypothesis,
a special experiment with two-storey emulsion chamber is currently
running at the Pamir, however, so far there are no results
available from it. The schematic view of expected results and its
comparison to our proposition is shown in Fig.1. The most
characteristic feature expected is the occurrence of a bump in the
absorption curve. It is caused by the fact that the chamber has an
air gap (2.5 m) where an effective decay of charmed particles
should take place. In the case of fluctuating cross section we
observe nonexponential behavior in the whole chamber. The second
hypothesis still awaits its possible future check.

\clearpage

\end{document}